\documentclass[lettersize,journal]{IEEEtran}
\usepackage{amsmath,amsfonts}
\usepackage{algorithmic}
\usepackage{algorithm}
\usepackage{array}
\usepackage{textcomp}
\usepackage{stfloats}
\usepackage{url}
\usepackage{verbatim}
\usepackage{graphicx}
\usepackage{cite}

\usepackage{algorithmic}

\usepackage{textcomp}
\usepackage{amsthm}
\newtheorem{theorem}{Theorem}
\newtheorem{lemma}{Lemma}
\usepackage{xcolor}
\def\BibTeX{{\rm B\kern-.05em{\sc i\kern-.025em b}\kern-.08em
    T\kern-.1667em\lower.7ex\hbox{E}\kern-.125emX}}
\graphicspath{{./images/}}

\usepackage[font=footnotesize]{caption}
\usepackage[font=scriptsize]{subcaption}
\begin{document}

\title{Probability of Error for Optimal Codes in a Reconfigurable Intelligent Surface Aided URLLC System\\
%for IEEE Journals and Transactions
}

% \author{\IEEEauthorblockN{Author 1}\\
% % \IEEEauthorblockA{\textit{School of Electrical and Information Engineering} \\
% % \textit{University of Sydney}\\
% % Sydney, Australia \\
% % likun.sui@sydney.edu.au}\\
% \and
% \IEEEauthorblockN{Author 2}\\
% % \IEEEauthorblockA{\textit{School of Electrical and Information Engineering} \\
% % \textit{University of Sydney}\\
% % Sydney, Australia \\
% % zihuai.lin@sydney.edu.au}
% }
\author{\IEEEauthorblockN{Likun Sui},\ \IEEEauthorblockN{Zihuai Lin}, ~\IEEEmembership{Senior Member,~IEEE,}
        % <-this % stops a space
\thanks{Likun Sui, and Zihuai Lin are with the \IEEEauthorblockA{\textit{School of Electrical and Information Engineering}}, The \textit{University of Sydney}, Sydney, NSW 2006, Australia (e-mail: likun.sui@sydney.edu.au; zihuai.lin@sydney.edu.au).}% <-this % stops a space
}



\maketitle

\begin{abstract}
The lower bound on the decoding error probability for the optimal code given a signal-to-noise ratio and a code rate are investigated in this letter for the reconfigurable intelligent surface (RIS) communication system over a Rician fading channel at the short blocklength regime, which is the key characteristic of ultra-reliable low-latency communications (URLLC) to meet the need for strict adherence to quality of service (QoS) requirements. Sphere packing technique is used to derive our main results. The Wald sequential t-test lemma and the Gaussian-Chebyshev quadrature are the main tools to obtain the closed-form expression for the lower bound. Numerical results are provided to validate our results and demonstrate the tightness of our results compared to the Polyanskiy-Poor-Verdu (PPV) bound.
\end{abstract}

\begin{IEEEkeywords}
Reconfigurable intelligent surfaces, ultra-reliable and low-latency communications, Lower bound, decoding error probability, sphere packing
\end{IEEEkeywords}

\section{Introduction}
\label{introduction}
\IEEEPARstart{N}{owadays}, industrial communications have differed significantly from conventional wireless communications as they necessitate deterministic communication with stringent quality of service requirements. These requirements include ultra-reliable and low-latency communications (URLLC), achieved by exchanging a small amount of data such as control commands or measurement data. For URLLC, it is crucial to maintain a maximum transmission latency of one millisecond, while ensuring the packet error probability falls within the range of $10^{-6}$ to $10^{-9}$\cite{1,URLLC}. To surmount the aforementioned obstacle, the reconfigurable intelligent surface (RIS) has emerged as a promising solution that has piqued significant research interest from both academia and industry \cite{2,3,4,WirelessCommLett_likun,RIS_likun,AmBC_likun}. To avoid the disconnection due to blockages by the obstacles, the RIS is regarded as a promising solution, which the RIS stands apart from conventional relay technologies by transforming the harsh propagation environment into a favorable one, due to its distinctive properties. These properties enhance the signal quality at the receiver side without requiring any additional power consumption and increase the diversity gain \cite{b10}. In contrast to existing relay technologies, the RIS can transform an unfavorable propagation environment into a favorable one due to its unique characteristics, which can improve signal quality at the receiver side without requiring more power. In addition to higher data rates, low latency is an essential factor that we must prioritize. Applications of the Internet of Things (IoT) \cite{leng2020implementation,IoT_FD,RF_energy1,IoTbook} and the fifth Generation (5G) communication system \cite{5G_ming,cellular11,5G_Yang1,5G_Yang2,MIMO_capacity,Chen2021MIMO,network_capacity} need to balance between low latency and high accuracy. An RIS is a panel, typically square or circular, composed of numerous reflecting elements. These elements induce independent phase shifts on the impinging electromagnetic (EM) waves. By meticulously designing the phase shifts of each element, the reflected EM waves can be constructively added to the direct signal from the base station (BS), but at most time the direct signal is blocked by an obstacle, increasing the signal power at the intended user and improving the system's signal-to-noise (SNR) performance. \par
As a result, the performance analysis for the RIS system at the short blocklength regime is much needed. Most of the existing contributions on the performance analysis in the short blocklength focus on the fundamental channel model, i.e., the additive white Gaussian noise (AWGN) channel. The authors in \cite{R1} identified Shannon's lower bound, specifically the sphere packing bound, as the superior choice for the AWGN channel, setting it as a benchmark for comparison against other bounds, such as the Gallager bound, Feinstein bound, and their own Polyanskiy-Poor-Verd\a'{u} (PPV) bound. In \cite{R2}, the authors demonstrated that the sphere packing bound provides an optimal lower bound for low-rate codes up to a specific blocklength, which can extend to several thousands. Additionally, the majority of existing works on RIS concentrate on the optimization of the active beamforming at the BS and the phase shift matrix at the RIS for different channel state information (CSI). In \cite{R3}, the authors derived the average decoding error probability using the PPV bound with a Gaussian input for RIS-aided communications.\par
\textbf{Contributions:} 1) This letter investigates the sphere packing bounds for evaluating the performance of an RIS-aided wireless system over a Rician fading channel at a short blocklength regime. 2) Using Wald sequential t-test lemma and the Gaussian-Chebyshev quadrature, we obtain a closed-form expression for the bound. 3) The approximation of the angle in the sphere packing technique used to determine the code rate in \cite{b6_1} and \cite{R2} will have a margin of error at a short blocklength regime, which we overcome by deriving the expression to calculate the exact value of the angle. 4) The results are validated through some numerical and Monte-Carlo simulation results and we set the PPV bound \cite{R1} as a reference.\par

\section{System Model}\label{system model}
\begin{figure}[t]
\centering
\includegraphics[width=6.5cm]{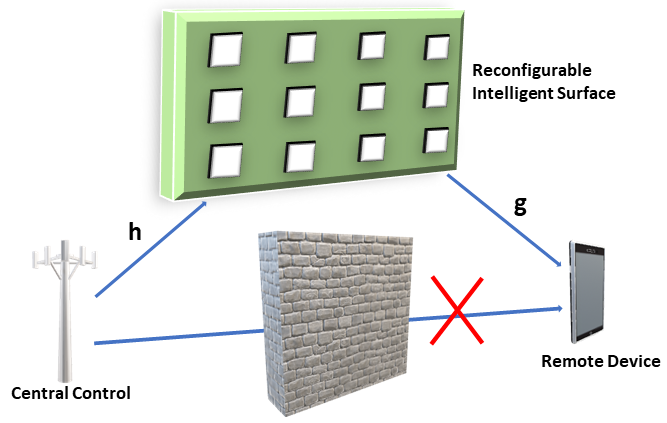}
\caption{System model for the RIS-aided wireless system}
\label{sys_model}
\vspace{-5mm}
\end{figure}

As shown in Fig. \ref{sys_model}, we consider an RIS-aided transmission system with the direct link blocked by an obstacle, i.e., a wall or building, between the single-element transmit and receive antennas. In this letter, we focus on enhancing the overall system performance by employing a rectangular RIS consisting of $N_{ris}$ elements. We assume that all the RIS elements are ideal which means that each of them can independently influence the phase, which can be flexibly adjusted, and the reflection angle of the impinging wave.\par
The signal vector at the receive antenna is given by
\begin{equation}
    y(i)=\sqrt{P_r} A(i) s(i)+w(i), \quad i=1,\dots,n,
\end{equation}
where $n$ and $i$ denote the blocklength and its $i$-th entry and $w(i)$ is the equivalent baseband AWGN with a zero mean and variance $N_{0}$, i.e., $w(i) \sim \mathcal{CN} (0, N_{0})$, and $A(i)$ is the the end-to-end channel coefficient, $P_r$ is the average power at the receiver, which can be obtained from the Friss path-loss model\cite{b11} as
\begin{equation}
    P_r = \frac{PG_tG_r\lambda^2}{16\pi^2(d_1+d_2)^2},
\end{equation}
where $P$ is signal power at the transmitter, and $\lambda$ is the wavelength in meters, and $G_t$ and $G_r$ are the transmitter and receiver antenna gains. $d_1$ and $d_2$ denote the distance between the transmitter and the RIS, the RIS and the receiver, respectively. We assume that the binary phase shift keying (BPSK) modulation is adopted in the air interface. The transmitted signal $s(i)\in\mathcal{A},\, i=1,\dots,n$, where $\mathcal{A}=\{-1,+1\}$.\par
% We assume that the total average transmit power has a maximum value of $P$, i.e., $\mathbb{E}\{\mathbf{X}^H\mathbf{X}\}\leq P$.
We denote the baseband equivalent channels between the transmitter and the $m$-th reflecting element of the RIS by $\mathbf{h}_{m}=[h_{m}(1),\dots,h_{m}(n)], \quad \forall m = 1,\dots,N_{ris}$ with $h_{m}(i)=|h_{m}(i)|e^{j\phi_{m}(i)}$, where $|h_{m}(i)|$ and $\phi_{m}(i)$ represent the amplitude and phase of the channel coefficient $h_{m}(i)$, respectively. The reflecting channels between the $m$-th reflecting element and the receiver are $\mathbf{g}_{m}=[g_{m}(1),\dots,g_{m}(n)]$ with $g_{m}(i)=|g_{m}(i)|e^{j\varphi_{m}(i)}$, where $|g_{m}(i)|$ and $\varphi_{m}(i)$ represent the amplitude and phase of the channel coefficient $g_{m}(i)$, respectively. The channels are assumed to be independent and identically distributed (i.i.d), and their envelops follow the Rician distribution, i.e., we have $|h_{m}(i)|\sim \textrm{Rician}(\zeta_1,\eta_1)$, where the shape parameter of the Rician fading $K_1=\eta_1^2/(2\zeta_1^2)$ denotes the ratio of the power contributions by a line-of-sight path to the remaining multipaths, and the scale parameter of the Rician fading $\Omega=2\zeta_1^2+\eta_1^2$ is the total power received in all paths. We assume that the probability density function (PDF) of $|g_m(i)|$ is also Rician distributed, i.e., $|g_m(i)|\sim \textrm{Rician}(\zeta_2,\eta_2)$. Then, $A(i)$ of our RIS-aided system can be expressed as
\begin{equation}\label{eqqq}
    A(i) = \sum_{m=1}^{N_{ris}}\eta_{m}(i)h_{m}(i)g_{m}(i),
\end{equation}
where $\eta_{m}(i)$ denotes the reflecting coefficient of the $m$-th reflecting element with $\eta_{m}(i)=|\eta_{m}(i)|e^{j\theta_{m}(i)}$, where $|\eta_{m}(i)|$ represents the reflecting gain and $\theta_{m}(i)$ is the phase shift configured by the $m$-th reflecting element. Without loss of generality, we assume the reflecting gain $|\eta_{m}(i)|=1$. In addition, we assume that the phases of the channels $h_{m}(i)$ and $g_{m}(i)$ are perfectly known to the transmitter, and that the RIS can choose the optimal phase shifting, i.e., $\theta_{m}(i)=-(\phi_{m}(i)+\varphi_{m}(i))$. Hence, (\ref{eqqq}) can be re-written as
\begin{equation}\label{eqqq_1}
    A(i) = \sum_{m=1}^{N_{ris}}|h_{m}(i)|\cdot |g_{m}(i)|
\end{equation}\par
\begin{figure}[t]
\centering
\includegraphics[width=5.0cm]{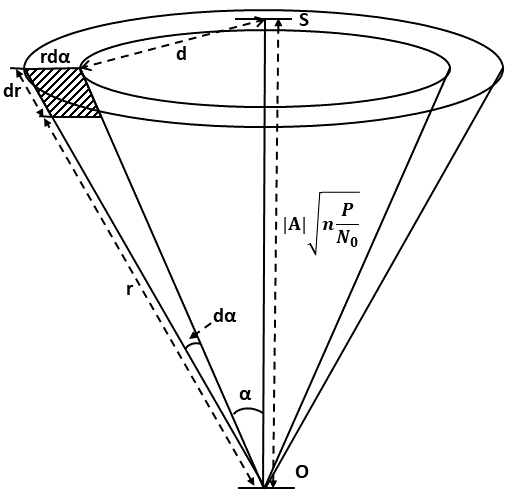}
\caption{Plane of cone of angle $\alpha$}
\label{sphere}
\vspace{-5mm}
\end{figure}
In Fig. \ref{sphere}, we assume that $O$ is the origin of an $n$-dimensional sphere and $S$ is a signal point which situates on the surface of the $n$ dimensional sphere. $\alpha$ denotes the angle of the cone intersected by the two outer lines and the angle $\alpha+d\alpha$ is the slightly larger angle which represents a larger cone. The radius of the outer cone is $r+dr$, and that of the inner cone is $r$. Therefore, the two sides of the ring-shaped plane are $rd\alpha$ and $dr$, respectively. $d$ presents the distance between the signal point $S$ and the above ring. We consider that there is a channel code with the number of codewords $M$. For notation simplicity, in the following, we use $A$ instead of $A(i)$, the code can place its $M$ points arbitrarily on the surface of the $n$-dimensional sphere whose radius is $|A|\sqrt{n\frac{P}{N_{0}}}$.\par
\section{Performance Analysis}\label{lower}
\subsection{The lower Bound}
\begin{theorem}\label{the}
For any $\alpha_1<\frac{\pi}{2}$, given the channel coefficient $A$, the conditional probability of error $P^{L}_{e,\ opt}(n,R|A)$ for the optimal code with the length $n$, and code rate $R$ can be lower-bounded by  
\begin{equation}
    P^{L}_{e,\ opt}(n,R|A)\geq\Phi(\alpha, n, A),
\end{equation}
where $\Phi(\alpha, n, A)$ denotes the probability of a signal point $S$ in the $n$-dimensional sphere, which has the $|A|\sqrt{n\frac{P}{N_{0}}}$ distance between the origin point $O$, being moved outside the cone whose angle is $\alpha$. And $\Lambda(\alpha, n)$ denotes the $(n-1)$-dimensional area of the cap, which is cut out by the cone on the $(n-1)$-dimensional unit sphere. And $\alpha_1$ is the angle which can even cut the sphere into $M$ parts, equivalently $M\Lambda(\alpha_1, n)=\Lambda(\pi, n)$. Then by taking the expectation over the channel coefficient $A$, the lower bound of the error probability $P^{L}_{e,\ opt}(n, R)$ for the optimal code with the length $n$, and code rate $R$ can be finally obtained as
\begin{equation}
    P^{L}_{e,\ opt}(n,R)\geq\mathbb{E}_{A}\Big[\Phi(\alpha, n, A)\Big].
\end{equation}
where the parameter $\alpha_{1}$ is determined by solving the equation
\begin{equation}\label{alpha}
    2^{nR}\Lambda(\alpha_1, n)=\Lambda(\pi, n).
\end{equation}
\end{theorem}\par
The proof of Theorem \ref{the} can be found in \cite{b6_1}\cite{R2}.

According to Theorem \ref{the}, at first, we need to calculate $\Lambda(\alpha_1, n)$. The surface of an $n$-dimensional sphere with an $r$ radius can be given as $S_n(r) = {n\pi^{n/2}r^{n-1}}/{\Gamma(\frac{n+2}{2})}$. According to the cosine law and the fact of the $(n-1)$-dimensional unit sphere, we obtain that the radius of the cap is $\sin{\alpha}$. Thus, we get 
\begin{equation}\label{1}
    \Lambda(\alpha_1, n) = \frac{(n-1)\pi^{\frac{n-1}{2}}}{\Gamma(\frac{n+1}{2})}\int_{0}^{\alpha_1}\sin^{n-2}{\alpha}d\alpha.
\end{equation}
From (\ref{1}), we can easily get
\begin{equation}\label{eq11}
    \Lambda(\pi, n) = S_n(1) = \frac{n\pi^{n/2}}{\Gamma(\frac{n+2}{2})}.
\end{equation}
In order to decrease the margin of error at the short blocklength regime and get the exact expression for $\Lambda(\alpha, n)$, we need to calculate $\int_{0}^{\alpha}(\sin{\alpha})^{n-2}d\alpha$ which is defined by the recursive relationship for all $n>3$
\begin{align*}
    &\Pi(\alpha, n) = \int_{0}^{\alpha}(\sin{\alpha})^{n-2}d\alpha
    \\&=-\sin^{n-3}{\alpha}\cos{\alpha}+(n-3)\int_{0}^{\alpha}\cos^2{\alpha}\sin^{n-4}{\alpha}d\alpha\\
    &= -\frac{\sin^{n-3}{\alpha}\cos{\alpha}}{n-2}+\frac{n-3}{n-2}\Pi(\alpha, n-2)
\end{align*}
and by the initial conditions $\Pi(\alpha, 2) \overset{\triangle}{=} \alpha$ and $\Pi(\alpha, 3) \overset{\triangle}{=} -\cos{\alpha}$.

For numerical accuracy purposes, we use this recursive relationship to compute the value of $\Pi(\alpha, n)$, and times the coefficient, then we have $\Lambda(\alpha,n)=\Pi(\alpha, n)(n-1)\pi^{\frac{n-1}{2}}/\Gamma(\frac{n+1}{2})$. When $n$ is even, we have
\begin{multline}\label{eq9}
\Lambda(\alpha, n)=\frac{(n-1)\pi^{\frac{n-1}{2}}}{\Gamma(\frac{n+1}{2})}\Big(\frac{(n-3)!!}{(n-2)!!}\alpha-\frac{\sin^{n-3}{\alpha}\cos{\alpha}}{n-2}\\
-\sum_{i=0}\sin^{n-5-2i}{\alpha}\cos{\alpha}\frac{\prod_{j=\frac{n}{2}-i}^{\frac{n}{2}}(2j-3)}{\prod_{k=\frac{n}{2}-(i+1)}^{\frac{n}{2}}(2k-2)}\Big)
\end{multline}
and when $n$ is odd, we have
\begin{multline}\label{eq10}
    \Lambda(\alpha, n)=\frac{(1-n)\pi^{\frac{n-1}{2}}}{\Gamma(\frac{n+1}{2})}\Big(\frac{(n-3)!!}{(n-2)!!}\cos{\alpha}+\frac{\sin^{n-3}{\alpha}\cos{\alpha}}{n-2}
\\+\sum_{i=0}\sin^{n-5-2i}{\alpha}\cos{\alpha}\frac{\prod_{j=\frac{n}{2}-i}^{\frac{n}{2}}(2j-3)}{\prod_{k=\frac{n}{2}-(i+1)}^{\frac{n}{2}}(2k-2)}\Big),
\end{multline}
where $(\cdot)!!$ denotes the double factorial. Then we need to calculate $\Phi(\alpha, n)$, which corresponds to the probability of signal point $S$ being carried outside the cone by the noise. The noise, of which the mean value and variance are zero and one, respectively, is generated by an $n$-dimensional Gaussian distribution function
\begin{equation}\label{equ2}
    f(d) = \frac{1}{(2\pi)^{n/2}}e^{-d^2/2}.
\end{equation}
Based on the cosine law from Fig. \ref{sphere}, we have the expression for $d$ which is shown below
\begin{equation}\label{equ3}
    d^2 = r^2+A^2n\frac{P}{N_{0}}-2rA\sqrt{n\frac{P}{N_{0}}}\cos{\alpha}.
\end{equation}
The differential volume of the ring-shaped region equals the shaded area, which is $rdrd\alpha$ times the surface of an $(n-1)$-dimensional sphere whose radius is $r\sin{\theta}$
\begin{align}
    dV &= rdrd\alpha S_{n-1}(r\sin{\alpha})\nonumber\\\label{equ4}
    &=rdrd\alpha\frac{(n-1)\pi^{\frac{n-1}{2}}(r\sin{\alpha})^{n-2}}{\Gamma(\frac{n+1}{2})}.
\end{align}
We multiply the PDF in (\ref{equ2}) and the differential volume in (\ref{equ4}) and then substitute $d$ from (\ref{equ3}). Therefore, we can get the expression for $\Phi(\alpha)$,
\begin{multline}\label{equ5}
    \Phi(\alpha, n, A) = \frac{(n-1)\exp(-\frac{1}{2}A^2n\frac{P}{N_{0}})}{2^{n/2}\pi^{1/2}\Gamma(\frac{n+1}{2})}\int_{\alpha_1}^{\pi/2}\sin^{n-2}{\alpha}\cdot\\\int_{0}^{\infty}r^{n-1}\exp(-\frac{1}{2}r^2+rA\sqrt{n\frac{P}{N_{0}}}\cos{\alpha})drd\alpha\\+Q(A\sqrt{n\frac{P}{N_{0}}}),
\end{multline}
where $Q(\cdot)$ is the Q function, $Q(x)=\frac{1}{\sqrt{2\pi}}\int_{x}^{\infty}\exp(-\frac{1}{2}t^2)dt$.\par
In order to simplify (\ref{equ5}) and reduce its complexity, our objective is to derive the closed form of $\Phi(\alpha, n, A)$ by eliminating integrals with respect to $r$ and $\alpha$. Initially, we can achieve this by leveraging the Wald sequential $t$-test lemma in \cite{b9_1} to eliminate the integral with respect to $r$ which is shown as follows
\begin{multline}
    \int_{0}^{\infty}r^{n-1}\exp(-\frac{1}{2}r^2+rA\sqrt{n\frac{P}{N_{0}}}\cos{\alpha})dr \\\approx \sqrt{2\pi} \Delta(\alpha, n) (\frac{\nabla(\alpha, n)}{e})^{n-1}\exp\big(\frac{1}{2}\nabla^2(\alpha, n, A)\big),
\end{multline}
where 
\begin{multline*}
    \nabla(\alpha, n, A) \overset{\triangle}{=} A\nabla(\alpha,n)=\\\sqrt{n}A(\frac{1}{2}\sqrt{\frac{P}{N_{0}}}\cos{\alpha}+\sqrt{\frac{P\cos^2{\alpha}}{4N_{0}}+\frac{n-1}{n(k_2-k^2_1)}})
\end{multline*}
and 
\begin{multline*}
  \Delta(\alpha, n) \overset{\triangle}{=} \frac{1}{2}\bigg(\Big(1+\frac{k_2-k^2_1}{4}\big(\sqrt{A^2\cos^2{\alpha}\frac{P}{N_{0}}+\frac{4}{k_2-k^2_1}}\\-A\cos{\alpha}\sqrt{\frac{P}{N_{0}}}\big)^2\Big)^{-1/2}+\sqrt{\frac{\nabla^2(\alpha, n)}{\nabla^2(\alpha, n)+\frac{n-1}{k_2-k^2_1}}}\bigg),
\end{multline*}
% \begin{multline*}
%   \Delta(\alpha, n, A) = \frac{1}{2}\bigg(\Big(1+\frac{1}{4}\big(\sqrt{A^2\cos^2{\alpha}\frac{P}{N_{0}}+4}\\-A\cos{\alpha}\sqrt{\frac{P}{N_{0}}}\big)^2\Big)^{-1/2}+\sqrt{\frac{\bar{r}^2(\alpha, n, A)}{\bar{r}^2(\alpha, n, A)+n-1}}\bigg)
% \end{multline*}
where $k_1=\mathbb{E}[A]=\frac{1}{4}\pi N_{ris}L_{1/2}(-K_1)L_{1/2}(-K_2)$ and $k_2=Var[A]=N_{ris}\big((1+K_1)(1+K_2)-\frac{1}{16}\pi^2L_{1/2}^2(-K_1)L_{1/2}^2(-K_2)\big)$ with $L_p(\cdot)$ denotes the Laguerre polynomial. Then, we have
\begin{multline}\label{equ5_1}
    \Phi(\alpha, n, A) \approx Q(A\sqrt{n\frac{P}{N_{0}}}) +\frac{(n-1)}{e^{n-1}2^{(n-1)/2}\Gamma(\frac{n+1}{2})}\\\int_{\alpha_1}^{\pi/2} \frac{\Delta(\alpha, n)}{\nabla(\alpha, n, A)\sin^2{\alpha}}  \exp\Big(-n\big(\frac{1}{2}\frac{A^2P}{N_{0}}-\frac{1}{2n}\nabla^2(\alpha, n, A)\\-\log(\nabla(\alpha,n, A)\sin{\alpha})\big)\Big)d\alpha.
\end{multline}

Then, to eliminate the integral with respect to $\alpha$, we utilize the Gaussian-Chebyshev quadrature. according to \cite{5}, the Gaussian-Chebyshev quadrature has been widely used in the state-of-the-art works since it approximates the integrals with limited terms to reduce the computational complexity while obtaining a good accuracy. Then, we have the closed-form expression for $\Phi(\alpha, n, A)$ as follows
\begin{multline}\label{19}
    \Phi(\alpha, n, A) \approx Q(A\sqrt{n\frac{P}{N_{0}}})+\frac{1}{e^{n-1}2^{(n-1)/2}\Gamma(\frac{n+1}{2})}\\\sum_{i=1}^{K} \frac{(\pi/2-\alpha_{1})w_{i}\sqrt{1-\phi_{i}}\Delta(s_{i}, n)}{2\nabla(s_{i}, n, A)\sin^2{s_{i}}}  \exp\Big(-n\big(\frac{1}{2}\frac{A^2P}{N_{0}}\\-\frac{1}{2n}\nabla^2(s_{i}, n, A)-\log\big(\nabla(s_{i},n, A)\sin{s_{i}}\big)\big)\Big),
\end{multline}
where 
\begin{equation}\label{s1}
    s_i=\frac{\pi/2-\alpha_1}{2}\psi_i+\frac{\pi/2+\alpha_1}{2}
\end{equation}
with $\psi_i=cos{(\frac{2i-1}{2K}\pi)}$ and $w_i=\frac{\pi}{K}$. 
Moreover, we need to get the probability density of the channel coefficient $A$ to calculate the expectation over $A$. The PDF of $A$ in (\ref{eqqq_1}) can be statistically evaluated as \cite{b10}
\begin{equation}\label{channel}
    f_A(x) = \frac{x^a}{b^{a+1}\Gamma(a+1)}\exp\big(-\frac{x}{b}\big),
\end{equation}
where 
\begin{equation}\label{a/b}
    a=\frac{k_1^2}{k_2}-1 \quad \textrm{and} \quad b=\frac{k_2}{k_1}.
\end{equation}
Finally, applying the following lemma, we combine (\ref{channel}) and (\ref{19}) together to obtain the closed-form expression for the lower bound denoted as $P_{e,\ opt}^L(n, R)$ in (\ref{eq6}), where $X(\alpha, n)={nP}/{(2N_{0})}-{\nabla^{2}(\alpha,n)}/{2}$, and $s_{i}$ is given in (\ref{s1}), and $a$, $b$ are given in (\ref{a/b}) with $P=\frac{16\pi^2(d_{1}+d_{2})^2P_{r}}{G_{t}G_{r}\lambda^2}$.
\begin{lemma}\label{lemma}
For the integral from negative infinity to infinity 
\begin{equation*}
    \int_{-\infty}^{\infty}A^aexp{(-cA^2-\frac{A}{b})}dA,
\end{equation*}
the closed-form expression is shown below
\begin{multline}
    \frac{1}{2}c^{-\frac{a+2}{2}}\big(\sqrt{c}\Gamma(\frac{1+a}{2}){}_{1}F_{1}(\frac{1+a}{2},\frac{1}{2},\frac{1}{4b^2c})\\-\frac{1}{b}\Gamma(1+\frac{a}{2}){}_{1}F_{1}(1+\frac{a}{2},\frac{3}{2},\frac{1}{4b^2c})\big),
\end{multline}
where ${}_1F_1(\cdot, \cdot, \cdot)$ denotes Kummer confluent hypergeometric function.
\end{lemma}
\begin{figure*}[t]
\begin{multline} \label{eq6}
P_{e,\ opt}^{L}(n, R)= \frac{1}{e^{n-1}2^{(n+1)/2}b^{a+n}\Gamma(a+n)\Gamma(\frac{n+1}{2})}\sum_{i=1}^{K} \frac{(\pi/2-\alpha_{1}w_{i}\sqrt{1-\phi_{i}})\Delta(s_{i}, n)}{\nabla^{-n+1}(s_{i}, n)\sin^{-n+2}(s_{i})} X^{-\frac{a+n+1}{2}}(s_{i}, n) \\\big(\sqrt{X(s_{i}, n)}\Gamma(\frac{a+n}{2}){}_1F_1(\frac{a+n}{2}, \frac{1}{2}, \frac{1}{4b^2X(s_{i}, n)})-\frac{1}{b}\Gamma(\frac{a+n+1}{2}){}_1F_1(\frac{a+n+1}{2}, \frac{1}{2}, \frac{1}{4b^2X(s_{i}, n)})\big)
\end{multline}
\hrulefill
\vspace{-5mm}
\end{figure*}
Compared with the approximation of $\alpha_{1}$ in \cite{b6_1}\cite{R2} which is tightly suitable when $R$ and $n$ grow relatively large. To compute the exact value of $\alpha_{1}$ in (\ref{alpha}) at a short blocklength regime, in the case of an even $n$, we use (\ref{eq9}) divided by (\ref{eq11}) to obtain the exact value of $\alpha_{1}$. Otherwise, we use (\ref{eq10}) divided by (\ref{eq11}) to calculate $\alpha_{1}$. 
\subsection{Asymptotic Formula}
At large enough blocklength, there exist some asymptotic formula to give a very good approximation of the bound and to allow all the computations in the logarithmic domain. We investigate the asymptotic formular of our derived result over the RIS system.
At first, we define $G(\alpha,A)\overset{\triangle}{=} AG(\alpha)=\frac{A}{2}(\sqrt{\frac{P}{N_0}}cos{\alpha}+\sqrt{\frac{P}{N_0}{cos}^2{\alpha}+\frac{4}{k_2-k_1^2}})$.
Then, the asymptotic formular for the lower bound is shown below.
\begin{multline}
    P_{e,opt,asymptotic}^L\geq\mathbb{E}_A\Bigg[\frac{\sqrt{n-1}}{6n(A\sqrt{\frac{P}{N_0}}+1)}\\exp\bigg\{-\frac{(A\sqrt{\frac{P}{N_0}}+1)^2+3}{2}-n\Big(\frac{A^2\frac{P}{N_0}}{2}-\frac{1}{2}G(\alpha_{1},A)A\\\sqrt{\frac{P}{N_0}}cos{\alpha_{1}}-log\big(G(\alpha_{1},A)sin{\alpha_{1}}\big)\Big)\bigg\}\Bigg].
\end{multline}
Moreover, we have the PDF of the channel coefficient A in (\ref{channel}). Thus, after applying Lemma \ref{lemma}, we have the closed-form expression of the asymptotic formula for the lower bound which is shown below
\begin{multline}
    P_{e,opt,asymptotic}^L\geq\frac{\sqrt{n-1}G^n(\alpha_1){sin}^n{\alpha_1}}{12e^2n\sqrt{\frac{P}{N_0}}\xi^{a+n}\Gamma(a+n)}\\X_2^{-\frac{a+n+1}{2}}(\alpha_1,n)\Big(\sqrt{X_2(\alpha_1,n)}\Gamma(\frac{a+n}{2})\\{}_{1}F_{1}\big(\frac{a+n}{2},\frac{1}{2},\frac{1}{4\xi^2X_2(\alpha_1,n)}\big)-\frac{1}{\xi}\Gamma(\frac{a+n+1}{2})\\{}_{1}F_{1}\big(\frac{a+n+1}{2},\frac{3}{2},\frac{1}{4\xi^2X_2(\alpha_1,n)}\big)\Big), 
\end{multline}
where $X_2\left(\alpha_1,n\right)=\frac{\left(n+1\right)P}{2N_0}-\frac{nG\left(\alpha_1\right)cos{\alpha_1}}{2}\sqrt{\frac{P}{N_0}}$ and $\xi=\frac{b}{1+b\sqrt{\frac{P}{N_0}}}$ with $P=\frac{16\pi^2(d_{1}+d_{2})^2P_{r}}{G_{t}G_{r}\lambda^2}$. 

\section{Comparison and Analysis}\label{Comparison and Analysis}
\begin{figure}[t]
     \centering
     \begin{subfigure}[b]{0.485\textwidth}
         \centering
         \includegraphics[width=\textwidth]{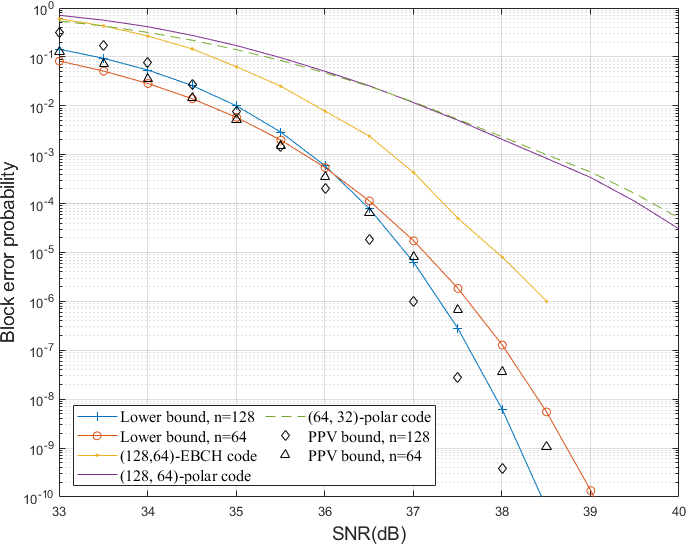}
         \caption{$N_{ris} = 4$}
         \label{comp_perf}
     \end{subfigure}
     \hfill
     \centering
     \begin{subfigure}[b]{0.485\textwidth}
         \centering
         \includegraphics[width=\textwidth]{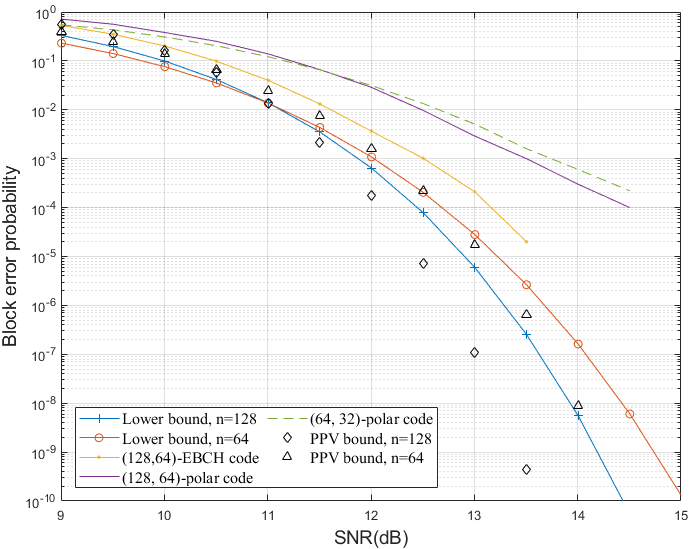}
         \caption{$N_{ris} = 64$}
         \label{comp1_perf}
     \end{subfigure}
     \caption{A comparison between the lower bound on the MLD decoding error probability for the codes of code length $n=64$ bits and $n=128$ bits with the same code rate $R=0.5$ bits per channel use over the perfect Rician fading channel in Section \ref{lower} for different number of RIS elements.}
        \label{comp_perf_two}
        \vspace{-5mm}
\end{figure}
In this section, we compare the lower bound for different blocklength and numbers of the RIS elements. Fig. \ref{comp_perf_two} illustrates the comparison between the lower bound on the MLD decoding error probability for the codes with the code length $n=128$ bits and $n=64$ bits and with the same code rate $R=0.5$ bits per channel use over the perfect Rician fading channel where the shape parameters $K_1$ and $K_2$ are chosen as $1$ and $0.5$ in Section \ref{lower} for different numbers of RIS elements $4$ and $64$, respectively. We define the transmit SNR as $\frac{P}{N_0}$ in decibels (dB) and set $P_r=0$dB, the wavelength $\lambda=0.125 m$ (the operating frequency $f_c=2.4$ GHz), $G_t=G_r=8$ ($9.03$ dBi) and $d_1=d_2=10 m$. We utilize the Polar code with the SCL decoder and the EBCH code with the OSD decoder to validate our bounds. All the simulations are averaged by $10^6$ Monte Carlo realizations. We set the PPV bound in \cite{R1} as a reference. From Fig. \ref{comp_perf_two}(\subref{comp_perf}), we observe that at the low SNR regime, the performance of the code with the short blocklength, i.e., $n=64$, is slightly better than the one with the long blocklength, i.e., $n=128$. When we increase the SNR, the long code will finally outperform the short code. The SNR's value of the intersection on the lower bounds is approximately $36$ dB. It indicates that the short code is preferred when the targeted SNR is less than $36$ dB. Otherwise, we can choose the long code to accomplish better transmission. In Fig. \ref{comp_perf_two}(\subref{comp_perf}), in terms of the code length of $128$ bits, for a decoding error probability of $10^{-2}$, the gap between the lower bound and the EBCH code is $0.8$ dB. When the decoding error probability level is low i.e., $10^{-4}$, the gap increases to $1.0$ dB.\par
%-7.75 dB --> 72.29 dB
Fig. \ref{comp_perf_two}(\subref{comp1_perf}) shows, as $N_{ris}$ increase from $4$ to $64$, the comparison between the performances of the lower bound of the different code length with the same code rate. When it comes to $n=128$ bits, for the decoding error probability of $10^{-2}$, the gap between its lower bound and the polar code is $0.7$ dB. It validates our derived result with different number of RIS elements. Furthermore, the SNR's value of the intersection on the lower bounds is approximately $11$ dB. In this work, we only simulated EBCH and polar codes, in the future work, we will also try different codes, such as rateless codes \cite{distributedRaptor,Raptor_ML,JNCC,fountaincodes2014} and convolutional codes \cite{RCRC,codedcpm1,codedcpm2,codedcpm3,codedcpm4,codedcpm6}.

\section{Conclusion}\label{conclusion}
This letter investigated the lower bound of the decoding error probability for the optimal code of the specific length, SNR and code rate for the RIS assisted URLLC communication system over a Rician fading channel at a short blocklength regime. The sphere packing technique is mainly used to derive our derived bound with the closed-form expression. Numerical and Monte-Carlo simulation results of EBCH and polar codes validate the derived results and demonstrate that the tightness of our results is better than the PPV bound at the high SNR regime.


\newpage

\vfill

\end{document}